\documentclass[slac_one]{revtex4}
\usepackage{graphicx}
\usepackage{fancyhdr}
\newcommand{\MET}{\mbox{$\raisebox{.3ex}{$\not\!$}E_T$}}
\newcommand{\MPT}{\mbox{$\raisebox{.3ex}{$\not\!$}p_T$}}
\begin{document}

\title{Searches at the Tevatron for a Low Mass Standard Model Higgs Boson} 

%

\author{B. Stelzer (on behalf of the CDF and D0 Collaborations)}
\affiliation{Simon Fraser University, BC, V5A 1S6, Canada}
\begin{abstract}
\vspace{-.5cm}
We report on searches for the standard model Higgs boson with the
CDF and D0 detectors using up to 2.7 fb$^{-1}$ of $\sqrt s=$~1.96 TeV
proton anti-proton collisions at Fermilab's Tevatron. 
We focus on analyses that are sensitive to low mass Higgs production 
with $m_H<$~140 GeV/$c^2$ favored by precision electroweak constraints.
Both CDF and D0 find no evidence for Higgs boson production, and set upper limits
on cross section times branching ratio.
Several analyses improvements have increased the sensitivity to a 
Higgs signal faster than what is expected from increasing datasets alone.
\end{abstract}
\maketitle
\thispagestyle{fancy}
\vspace{-1.cm}
\section{INTRODUCTION} 
The standard model of particle physics (SM) incorporates electroweak
gauge symmetry breaking via the Higgs mechanism \cite{higgs}. The resulting
Higgs boson of the theory has yet to be observed experimentally.
The Higgs mass is not predicted by the theory but can be constrained
due to its predicted couplings to other particles.
Global fits to precision electroweak data have been updated
to incorporate the new world average of the 
$W$ boson mass, $m_W = 80.398\pm0.025$ GeV/$c^2$, and the top quark mass,
$m_t = 172.4\pm1.2$ GeV/$c^2$ \cite{mwtop}. The fit favors a light
Higgs boson with mass $m_H=84^{+34}_{-26}$~GeV/$c^2$, which translates into an upper
limit of 154 GeV/$c^2$ at 95\% confidence level (C.L.)~\cite{ewkfit}.
Results from direct searches at the LEP experiments exclude Higgs mass
values lower than 114.4 GeV/$c^2$~\cite{lepHiggs}. This motivates searches for a Higgs boson 
just above the limits set by LEP. Both Tevatron experiments, CDF and D0,
recognize the search for the SM Higgs boson as one of their highest priorities.

For high Higgs boson masses, with  $m_H>$~140 GeV/$c^2$, the decay $H\rightarrow W^{+}W^{-}$
dominates. For low Higgs boson masses with  $m_H<$~140 GeV/$c^2$, the decay
$H\rightarrow b\bar{b}$ dominates, with a branching ratio (BR) of 73 \%
at $m_H=115$~GeV/$c^2$~\cite{Hxsec}. Even though Higgs production 
through gluon fusion $gg\rightarrow H\rightarrow b\bar{b}$ has the largest
cross section times BR~\cite{Hxsec}, the data sample with bottom quark jets is dominated by non-resonant
QCD multi-jet background. As a result, searches for production of a Higgs boson
in association with a $W$ or a $Z$ boson yield the most sensitive
results because the background rate is reduced by demanding the presence
of a reconstructed $W$ or $Z$ boson.
\subsection{Bottom Quark Identification}
Efficient $b$ quark identification with high purity is a key element when
searching for $H\rightarrow b\bar{b}$ signatures. Jets originating from $b$
quarks can be ``tagged'' by reconstructing the $b$ decay vertex, displaced
from the primary interaction vertex due to the relatively long $b$ lifetime.
CDF employs secondary vertex and jet lifetime tagging algorithms to select $b$ jets.
The purity is improved in a second step using a Neural Network jet-flavor separator. 
D0 uses a Neural Network $b$ tagging algorithm based on secondary
vertex and jet lifetime tagging information to 
select candidate jets at specific operating points
of the Neural Network output. Using this tool, $b$ tagging efficiencies
have been improved by 33\% while keeping the rate of falsely identified
light flavor jets (mistags) low.
The efficiencies range between 40-70\% for $b$ jets 
at a low mistag rates between 0.5-3\% for light flavor jets.
\subsection{Advanced Analysis Techniques}
Almost all Higgs searches at the Tevatron employ advanced analysis techniques
like artificial Neural Networks (NN), boosted decision trees (BDT) or matrix
element techniques (ME) to combine kinematic characteristics of signal and
background events into a single discriminant. These techniques improve
the separation of signal to background over the invariant Higgs boson mass distribution
which is the most important single variable.
Careful validation of all input variables is mandatory for robust results.
\section{ANALYSES}
\subsection{Higgs Search in the $W^{\pm}H\rightarrow \ell^{\pm}\nu b\bar{b}$ Channel}
\begin{figure*}[!h]
\centering
\includegraphics[width=0.85\textwidth]{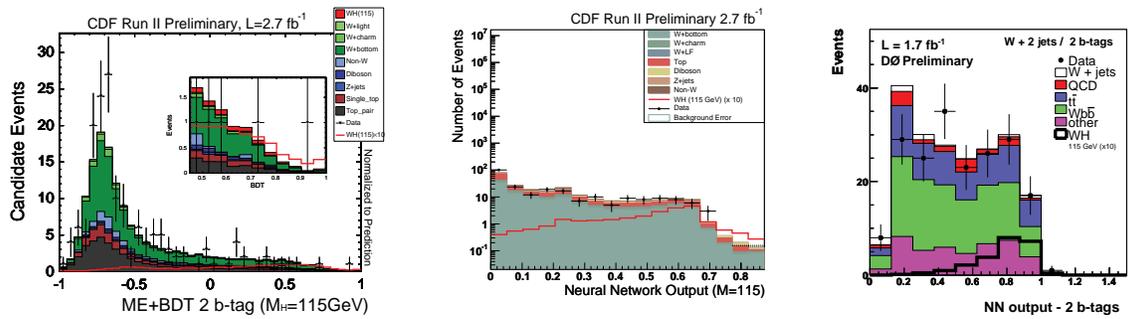}
\caption{Discriminant distributions used in the $W^{\pm}H\rightarrow \ell^{\pm}\nu b\bar{b}$ search
showing the two $b$ tag CDF ME+BDT (left), CDF NN (middle) and D0 NN (right) analysis.} 
\label{WHplots}
\end{figure*}

The experimental signature in this search channel
is an energetic lepton ($e^{\pm},\mu^{\pm}$), large missing transverse energy (\MET) from the neutrino
and two energetic jets of which one or two are $b$ tagged. 
The dominant background arise from $W+b\bar{b}$ and top production.
Several improvements have advanced the sensitivity in this channel.
Increased acceptance is obtained by extending the muon coverage, use of
track-only leptons, looser $b$ tagging and jet flavor purification. 
About 3-4 Higgs candidate events are expected per fb$^{-1}$ for $m_H =$ 115 GeV/$c^2$.
The sample is split into one and two $b$ tag events to improve sensitivity.
Both experiments use a NN discriminant and CDF recently added a combined ME+BDT technique.
Figure \ref{WHplots} shows the D0 NN, CDF NN and CDF ME+BDT analysis for two
$b$ tag candidate events, respectively. In the absence of a significant signal
the following upper limits at 95\% C.L. on the $WH$ production cross section times $H\rightarrow b\bar{b}$
BR are observed (expected) for $m_H =$ 115 GeV/$c^2$:
CDF NN: 5.0 (5.8), CDF ME+BDT: 5.8 (5.6), D0 NN 9.3 (8.5), expressed
in multiples of the SM prediction. D0 newly added a dedicated search
for $W^{\pm}H\rightarrow \tau^{\pm}\nu b\bar{b}$ with hadronic $\tau$ decays.
Using the di-jet mass distribution to separate signal from background a limit
of 35.4 (42.1) $\times$ SM is obtained in that channel.
\vspace{-.6cm}
\subsection{Higgs Search in the $ZH\rightarrow \ell^{+}\ell^{-}b\bar{b}$ Channel}
The $ZH\rightarrow l^{+}l^{-}b\bar{b}$ channel 
features a pair of energetic leptons and a pair
\begin{figure*}[b!]
\centering
\includegraphics[width=1.02\textwidth]{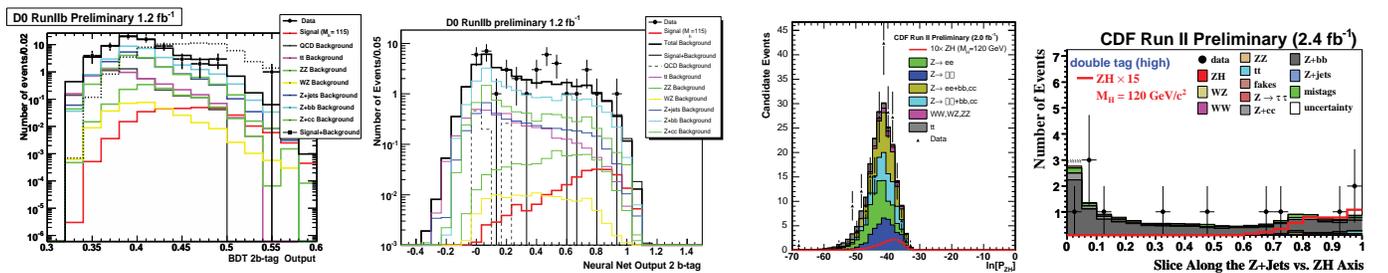}
\caption{
Discriminant distributions used in the $ZH\rightarrow \ell^{+}\ell^{-} b\bar{b}$ search
showing the two $b$ tag D0 NN and BDT used in the di-electron and di-muon channel (left), 
CDF ME (middle) and a slice of the CDF 2D NN analysis (right).} 
\label{ZHplots}
\end{figure*}
of energetic $b$ jets which allows to select a sample with very little non-$Z$ background. 
One Higgs event is expected per fb$^{-1}$ 
for $m_H =$ 115 GeV/$c^2$. Highlights in this analysis have been
the development of looser lepton identification to
increase the acceptance by making use of complementary
$Z$ boson triggers. The di-jet system is corrected with
a Neural Network which assigns \MET~to the jets according
to their \MET~projections and event topology.
The CDF analysis uses a two dimensional 
Neural Network discriminant based on nine 
kinematic variables in order to maximize 
discrimination of the signal from $Z$+jets
and top background. Another CDF analysis uses a matrix element
technique to search for a $ZH$ signal. The D0 analysis uses 
ten kinematic variables combined into a NN to separate
signal from background in the di-electron channel and a BDT
technique in the di-muon channel. Figure \ref{ZHplots} shows 
discriminant distributions from CDF and D0. The observed (expected) limits are: 
CDF 2D NN: 11.6 (11.8) $\times$ SM, D0 NN/BDT: 11.0 (12.3) $\times$ SM for 
$m_H =$ 115 GeV/$c^2$ and
CDF ME: 14.2 (15.0) $\times$ SM at $m_H =$ 120 GeV/$c^2$.
\subsection{Higgs Search in the $ZH\rightarrow \nu\bar{\nu} b\bar{b}$ / $W^{\pm}H\rightarrow (\ell^{\pm})\nu b\bar{b}$ Channel}
The $ZH\rightarrow \nu\bar{\nu} b\bar{b}$ signature features two energetic jets,
with one or two $b$ tags, recoiling against a $Z$ boson which decays
\begin{figure*}[h!]
\centering
\includegraphics[width=1.\textwidth]{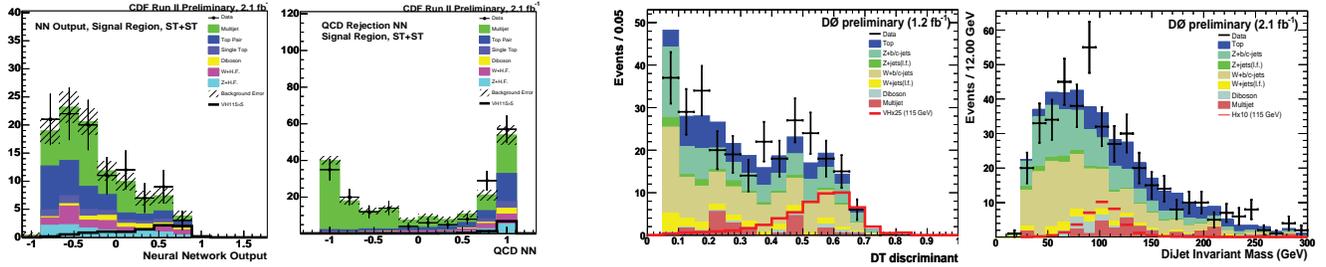}
\caption{Discriminant distributions used in the $ZH\rightarrow \nu\bar{\nu} b\bar{b}$ / $W^{\pm}H\rightarrow (\ell^{\pm})\nu b\bar{b}$ search
showing the two $b$ tag CDF NN and the QCD rejection NN (left) and the D0 BDT and di-jet mass distribution (right).} 
\label{fig:zh}
\end{figure*}
invisible into a pair of neutrino, producing large \MET.
Events from $W^{\pm}H\rightarrow \ell^{\pm}\nu b\bar{b}$ are also accepted
when the electron or muon escapes detection. Next to two jet events, also three jet events are accepted and capture cases
where a tau or a hard gluon radiation is reconstructed as an additional jet. Highlights in this analysis have been the implementation
of looser jet selection, extended $b$ tagging and improved di-jet mass resolution due
to track-based jet energy corrections.
About 3-4 Higgs candidate events are expected per fb$^{-1}$ for $m_H =$ 115 GeV/$c^2$.
Both, CDF and D0 employ a \MET~+ 2 Jets trigger to select candidate events.
The dominant background arises from QCD multi-jet production where mismeasured jets
lead to fake \MET. Both experiments reduce this background by imposing clever reconstruction
requirements like the alignment of calorimeter based \MET~and tracking based \MPT.
CDF developed a dedicated NN to reject QCD. Both experiments model the residual background from data. 
The CDF limit for $m_H =$ 115 GeV/$c^2$ 
is 7.9 (6.3) $\times$ SM based on a NN and the D0 limit is: 7.5 (8.4) $\times$ SM by means of a BDT analysis.
\subsection{Higgs Search in Complementary Channels}
Both experiments have made progress in exploring the prospects of
Higgs search channels which, by themselves, are not discovery
channels but add sensitivity in combination.
In the SM, the $H\rightarrow \gamma \gamma$ BR is
\begin{figure*}[b]
\centering
\includegraphics[width=1.0\textwidth]{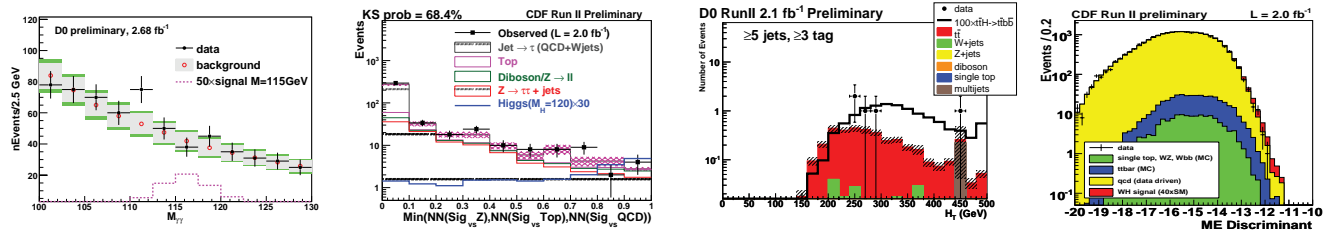}
\caption{From left to right: data and Monte Carlo prediction for the $H\rightarrow \gamma \gamma$ analysis,
the CDF $H\rightarrow \tau \tau$ NN analysis the D0 $t\bar{t}H$ analysis and the CDF $WH\rightarrow q\bar{q}b\bar{b}$ ME analysis.} 
\label{fig:comp}
\end{figure*}
only 0.22\% for a Higgs boson mass of 130 GeV/$c^2$. However,
in some fermiophobic models beyond the SM this BR 
can be significantly enhanced.  
D0 searches for an excess in the di-photon mass spectrum and 
sets an upper limit of 30.8 (23.2) $\times$ SM for $m_H =$ 115 GeV/$c^2$.
Similarly, the BR of $H\rightarrow \tau \tau$ is only about 10\% that of 
$H\rightarrow b\bar{b}$. CDF performs a search in $\tau^{+}\tau^{-}$ + 2 jets events
implemented as a simultaneous search for $WH$, $ZH$, vector boson fusion and gluon
fusion Higgs production. The limits obtained are 30.5 (24.8) $\times$ SM for $m_H =$ 115 GeV/$c^2$
using a NN technique.
Higgs production in association with top pairs, $t\bar{t}H$ is interesting,
because it tests the top Yukawa. The cross section at the Tevatron is tiny. D0 searches in lepton+jets
events with 4 and 5 or more jets and 2 or 3 $b$ tags for deviations in the distribution
of the total transverse energy in the event. The limit obtained is 
63.9 (45.3) $\times$ SM for $m_H =$ 115 GeV/$c^2$.
CDF has also performed a search in the all hadronic decay channel of
associated $WH$ and $ZH$ production. The analysis requires four energetic jets with at least
two $b$ tags. A data driven
background modeling has been developed to describe the overwhelming
background from QCD multi-jets. CDF sets upper limit of 37.0 (36.6) $\times$ SM for $m_H =$ 115 GeV/$c^2$ based on a ME 
discriminant technique.
\begin{table}[h!]
\begin{center}
\caption{Summary of all results for $m_H =$ 115 GeV/$c^2$.}
\begin{tabular}{l|cccc}
\hline 
Channel 					& \multicolumn{2}{c}{CDF 95\% C.L. limits} 		& \multicolumn{2}{c}{D0 95\% C.L. limits}\\
\hline 
$WH\rightarrow \ell\nu b\bar{b}$ (ME+BDT)		& 5.8 (5.6)	& {\it $\cal{L}=$2.7 fb$^{-1}$}	&		&  				\\
$WH\rightarrow \ell\nu b\bar{b}$ (NN)		& 5.0 (5.8) 	& {\it $\cal{L}=$2.7 fb$^{-1}$}	& 9.3 (8.5) 	& {\it $\cal{L}=$1.7 fb$^{-1}$} 	\\
$WH\rightarrow \tau\nu b\bar{b}$ (NN) ($m_{jj}$)		& 		&			& 35.4 (42.1) 	& {\it $\cal{L}=$0.9 fb$^{-1}$} 	\\
$ZH\rightarrow \nu\bar{\nu} b\bar{b}$ / $WH\rightarrow (\ell)\nu b\bar{b}$ (NN, BDT)	& 7.9 (6.3) 	& {\it $\cal{L}=$2.1 fb$^{-1}$}	& 7.5 (8.4) 	& {\it $\cal{L}=$2.1 fb$^{-1}$} 	\\
$WH,ZH\rightarrow q\bar{q} b\bar{b}$ (ME)		& 37.0 (36.6) 	& {\it $\cal{L}=$2.0 fb$^{-1}$}&				 	\\
$ZH\rightarrow \ell\ell b\bar{b}$ (NN, BDT)		& 11.6 (11.8) 	& {\it $\cal{L}=$2.4 fb$^{-1}$}& 11.0 (12.3) 	& {\it $\cal{L}=$2.3 fb$^{-1}$} 	\\
$ZH\rightarrow \ell\ell b\bar{b}$ (ME)		& 14.2 (15.0) 	& {\it $\cal{L}=$2.0 fb$^{-1}$}& 			&		\\
$t\bar{t}H\rightarrow \ell\nu b\bar{b}b\bar{b}q\bar{q}$ ($H_T$)& 		&			& 63.9 (45.3) 	& {\it $\cal{L}=$2.1 fb$^{-1}$} 	\\
$H\rightarrow \gamma \gamma$ ($m_{\gamma \gamma}$)			& 		&			& 30.8 (23.2) 	& {\it $\cal{L}=$2.7 fb$^{-1}$}	\\
$H\rightarrow \tau \tau$ ($m_{\tau \tau}$)			& 30.5 (24.8) 	& {\it $\cal{L}=$2.2 fb$^{-1}$}	&		&			\\
\hline 
\end{tabular}
\label{tab:summary}
\end{center}
\end{table}
\vspace{-1.cm}
\section{CONCLUSIONS}
Higgs physics at the Tevatron is entering an exciting phase.
The prospects for evidence or exclusion at the Tevatron depend
on large integrated luminosity, improved acceptance and
advanced analysis techniques to extract the small Higgs signal
from a large amount of background. 
The combination of Tevatron results is currently in progress.
The combined expected cross section limit for a Higgs mass of 115 GeV/$c^2$ should fall below
three times the SM prediction. Over the last year, creative analyses improvements have 
led to an increased sensitivity to a Higgs signal faster than
what is expected from the larger datasets alone \cite{cdfresults,d0results}.
\begin{acknowledgments}
I would like to thank my colleagues from the CDF and D0 collaborations for their dedicated work
in producing first class results. Compliments to the conference organizers for a terrific event.
I would like to acknowledge the A. v. Humboldt Foundation for supporting this research.
\end{acknowledgments}

\end{document}